\documentclass{PoS}

\usepackage{axodraw}
\usepackage{epsfig}

\def\nl{\nonumber\\}

\newcommand{\lsim}
{\mathrel{\raisebox{-.3em}{$\stackrel{\displaystyle <}{\sim}$}}}

\def\asymp#1%
{\mathrel{\raisebox{-.4em}{$\widetilde{\scriptstyle #1}$}}}

\def\Nequal#1%
{\mathrel{\raisebox{-.5em}{$\stackrel{=}{\scriptstyle\rm#1}$}}}
\newcommand{\dsl}[1]{\not \hspace{-0.7mm}#1}
\def\dsl{\mathpalette\make@slash}
\def\make@slash#1#2{\setbox\z@\hbox{$#1#2$}%
  \hbox to 0pt{\hss$#1/$\hss\kern-\wd0}\box0}

\def\beq{\begin{equation}}
\def\eeq{\end{equation}}
\def\bit{\begin{itemize}}
\def\eit{\end{itemize}}
\def\beqar{\begin{eqnarray}}
\def\eeqar{\end{eqnarray}}
\def\barr#1{\begin{array}{#1}}
\def\earr{\end{array}}
\def\bfi{\begin{figure}}
\def\efi{\end{figure}}
\def\btab{\begin{table}}
\def\etab{\end{table}}
\def\bce{\begin{center}}
\def\ece{\end{center}}
\def\nn{\nonumber}

\def\text{\textstyle}

\def\refeq#1{\mbox{(\ref{#1})}}

\def\reffi#1{\mbox{Fig.~\ref{#1}}}

\def\citere#1{\mbox{~\cite{#1}}}
\def\citeres#1{\mbox{~\cite{#1}}}

\newcommand{\TeV}{\unskip\,\mathrm{TeV}}
\newcommand{\GeV}{\unskip\,\mathrm{GeV}}
\newcommand{\MeV}{\unskip\,\mathrm{MeV}}

\newcommand{\fb}{\unskip\,\mathrm{fb}}

\newcommand{\ri}{{\mathrm{i}}}

\newcommand{\rd}{{\mathrm{d}}}

\newcommand{\rT}{{\mathrm{T}}}

\newcommand{\A}{{\cal{A}}}

\newcommand{\M}{{\cal{M}}}
\def\mathswitchr#1{\relax\ifmmode{\mathrm{#1}}\else$\mathrm{#1}$\fi}

\newcommand{\PW}{\mathswitchr W}

\newcommand{\PZ}{\mathswitchr Z}

\newcommand{\Pg}{\mathswitchr g}

\newcommand{\PH}{\mathswitchr H}

\newcommand{\Pb}{\mathswitchr b}

\newcommand{\Pp}{\mathswitchr p}
\newcommand{\Pj}{\mathswitchr j}
\newcommand{\Pt}{\mathswitchr t}

\def\mathswitch#1{\relax\ifmmode#1\else$#1$\fi}

\newcommand{\MH}{\mathswitch {M_\PH}}

\newcommand{\Mt}{\mathswitch {m_\Pt}}

\def\solid{\raise.9mm\hbox{\protect\rule{1.1cm}{.2mm}}}
\def\dash{\raise.9mm\hbox{\protect\rule{2mm}{.2mm}}\hspace*{1mm}}

\newcommand{\LO}{{\mathrm{LO}}}

\newcommand{\onel}{{\mbox{\scriptsize 1-loop}}}

\def\lra{\mathop{\mathrm{\leftrightarrow}}\nolimits}

\def\draftdate{\relax}
\def\mda{\relax}
\def\mua{\relax}
\def\mla{\relax}
\def\Mda{\relax}
\def\Mua{\relax}
\def\Mla{\relax}
\def\draft{
\def\thtystars{******************************}
\def\sixtystars{\thtystars\thtystars}
\typeout{}
\typeout{\sixtystars**}
\typeout{* Draft mode!
         For final version remove \protect\draft\space in source file *}
\typeout{\sixtystars**}
\typeout{}
\def\draftdate{\today}
\def\mua{\marginpar[\boldmath\hfil$\uparrow$]%
                   {\boldmath$\uparrow$\hfil}%
                    \typeout{marginpar: $\uparrow$}\ignorespaces}
\def\mda{\marginpar[\boldmath\hfil$\downarrow$]%
                   {\boldmath$\downarrow$\hfil}%
                    \typeout{marginpar: $\downarrow$}\ignorespaces}
\def\mla{\marginpar[\boldmath\hfil$\rightarrow$]%
                   {\boldmath$\leftarrow $\hfil}%
                    \typeout{marginpar: $\lra$}\ignorespaces}
\def\Mua{\marginpar[\boldmath\hfil$\Uparrow$]%
                   {\boldmath$\Uparrow$\hfil}%
                    \typeout{marginpar: $\uparrow$}\ignorespaces}
\def\Mda{\marginpar[\boldmath\hfil$\Downarrow$]%
                   {\boldmath$\Downarrow$\hfil}%
                    \typeout{marginpar: $\downarrow$}\ignorespaces}
\def\Mla{\marginpar[\boldmath\hfil$\Rightarrow$]%
                   {\boldmath$\Leftarrow $\hfil}%
                    \typeout{marginpar: $\lra$}\ignorespaces}
\overfullrule 5pt
\oddsidemargin -15mm
\marginparwidth 29mm
}

\def\stars{\strut\leaders\hbox{*}\hfill\strut}
\def\starline{\hfil\strut\hfil\hbox to \textwidth {\stars}\hfil}

\title{NLO QCD corrections to $\mathbf{\Pt\bar\Pt\Pb\bar\Pb}$ production at the LHC}

\ShortTitle{NLO QCD corrections to $\mathbf{\Pt\bar\Pt\Pb\bar\Pb}$ production at the LHC}

\author{Axel Bredenstein\\
High Energy Accelerator Research Organization (KEK),\\
Tsukuba, Ibaraki 305-0801, Japan\\
E-mail: \email{Axel.Bredenstein@gmx.de}}

\author{Ansgar Denner\\
Paul Scherrer Institut, W\"urenlingen und Villigen,\\
CH-5232 Villigen PSI, Switzerland\\
E-mail: \email{ansgar.denner@psi.ch}}

\author{Stefan Dittmaier\\
Albert-Ludwigs-Universit\"at Freiburg, Physikalisches Institut,\\
D-79104 Freiburg, Germany\\
E-mail: \email{stefan.dittmaier@physik.uni-freiburg.de}}

\author{\speaker{Stefano Pozzorini}%
\\
CERN, Theory Division\\
CH-1211 GENEVA 23\\      
E-mail: \email{Stefano.Pozzorini@cern.ch}}

\abstract{
The next-to-leading order (NLO) QCD corrections
to $\Pp\Pp\to\Pt\bar\Pt\Pb\bar\Pb$ at the LHC
reveal that the scale choice adopted in previous lowest-order 
simulations 
underestimates the $\Pt\bar\Pt\Pb\bar\Pb$
cross section by a factor two.
We discuss a new dynamical scale that 
stabilizes the perturbative predictions
and describe the impact of the corrections on 
the shape of distributions.
We also account for the techniques employed 
to compute the six-particle
one-loop amplitudes with high CPU efficiency.
}% end of abstract

\FullConference{RADCOR 2009 - 9th International Symposium on Radiative
Corrections
(Applications of Quantum Field Theory to Phenomenology) \\
        October 25-30 2009\\
        Ascona, Switzerland}

\begin{document}

\section{Introduction}

The discovery of the Higgs boson and the measurement of its interactions
with massive quarks and vector bosons represent a central goal of the 
%\cite{atlas-cms-tdrs,Aad:2009wy} 
%\cite{Ball:2007zza} 
Large Hadron Collider (LHC).
In the mass range $114\GeV<\MH\lsim 130 \GeV$, associated 
$\Pt\bar\Pt\PH$ production provides the opportunity to observe the
Higgs boson in the $\PH\to\Pb\bar\Pb$ decay channel and to
measure the top-quark Yukawa coupling. 
However, the extraction of the $\Pt\bar\Pt\PH(\PH\to\Pb\bar\Pb)$ signal 
from its large QCD backgrounds,
$\Pp\Pp\to\Pt\bar\Pt\Pb\bar\Pb$ and $\Pt\bar\Pt\Pj\Pj$,
represents a serious challenge.
The selection strategies elaborated by ATLAS and CMS~\cite{Aad:2009wy,Ball:2007zza},
which assume $30\fb^{-1}$  and  $60\fb^{-1}$, respectively,
anticipate a statistical significance 
around $2\sigma$ (ignoring systematic uncertainties)
and a signal-to-background ratio as low as 1/10.
This calls for better than 10\% precision in the
background description, a very demanding requirement both from the
experimental and theoretical point of view.
Very recently, a novel selection strategy
based on highly boosted Higgs bosons
%, which decay into ``fat jets'' containing two $\Pb$~quarks, 
has opened new
and very promising perspectives~\cite{Plehn:2009rk}.
This approach 
might %enable a better background suppression and 
increase the signal-to-background ratio beyond $1/3$. 

The calculation of the NLO QCD corrections to the irreducible 
$\Pt\bar\Pt\Pb\bar\Pb$ background, first presented in~\citeres{Bredenstein:2008zb,
Bredenstein:2009aj} and subsequently confirmed in~\citere{Bevilacqua:2009zn}, constitutes another important step towards the
observability of $\Pt\bar\Pt\PH(\PH\to\Pb\bar\Pb)$ at the LHC.
These NLO predictions are mandatory in order to reduce the huge
scale uncertainty of the lowest-order (LO) $\Pt\bar\Pt\Pb\bar\Pb$ cross
section, 
which can vary
up to a factor four if the QCD scales are
identified with different kinematic parameters~\cite{Kersevan:2002vu}.
Previous results for five-particle processes that feature a signature
similar to $\Pt\bar\Pt\Pb\bar\Pb$ indicate that setting the renormalization
and factorization scales equal to half the threshold energy,
$\mu_\mathrm{R,F}=E_{\mathrm{thr}}/2$, is a reasonable scale
choice.  At this scale the NLO QCD corrections to $\Pp\Pp\to\Pt\bar\Pt\PH$
($K\simeq 1.2$)~\cite{Beenakker:2001rj},
$\Pp\Pp\to\Pt\bar\Pt\Pj$ ($K\simeq$1.1)~\cite{Dittmaier:2007wz}, and
$\Pp\Pp\to\Pt\bar\Pt\PZ$ ($K\simeq 1.35$)~\cite{Lazopoulos:2008de}, are
fairly moderate. This motivated experimental groups to adopt the
scale $\mu_\mathrm{R,F}=E_{\mathrm{thr}}/2=\Mt+m_{\Pb\bar\Pb}/2$ 
for the LO simulation of the $\Pt\bar\Pt\Pb\bar\Pb$ background~\cite{Aad:2009wy}.
However, at this scale the NLO corrections to $\Pp\Pp\to\Pt\bar\Pt\Pb\bar\Pb$
turn out to be unexpectedly
large $(K\simeq 1.8)$~\cite{Bredenstein:2009aj,Bevilacqua:2009zn}.
As we argue, a reliable perturbative description of 
$\Pt\bar\Pt\Pb\bar\Pb$ production requires a different scale choice~\cite{Bredenstein:2010rs}.

The calculation of the NLO
corrections to $\Pp\Pp\to\Pt\bar\Pt\Pb\bar\Pb$ constitutes also 
an important technical benchmark. 
The description of many-particle processes at NLO plays 
an central role for the LHC physics programme,
and the technical challenges raised by
these calculations have triggered an impressive amount of 
conceptual and technical developments.
Within the last few months, this progress has lead to  
the first NLO results for six-particle processes 
at the LHC, namely for 
\mbox{$\Pp\Pp\to\Pt\bar\Pt\Pb\bar\Pb$}~\cite{Bredenstein:2009aj,Bevilacqua:2009zn},
the leading-~\cite{KeithEllis:2009bu} and the full-colour contributions~\cite{Berger:2009ep} to $\Pp\Pp\to\PW \Pj\Pj\Pj$, 
and for the $q\bar q$ contribution to 
$\Pp\Pp\to\Pb\bar\Pb\Pb\bar\Pb$~\cite{Binoth:2009rv}.

To compute the virtual corrections to $\Pt\bar\Pt\Pb\bar\Pb$
production we employ explicit diagrammatic representations of the
one-loop amplitudes and numerical reduction of 
tensor integrals~\cite{Denner:2002ii,Denner:2005nn}.
The factorization of colour matrices, 
the algebraic reduction of helicity structures,
and the systematic recycling of a multitude of common
subexpressions---both inside individual diagrams and in
tensor integrals of different diagrams that share common 
sub-topologies---strongly mitigate the factorial complexity 
that is inherent in Feynman diagrams and
lead to a remarkably high CPU efficiency.
The real corrections are handled with the dipole
subtraction 
method~\cite{Catani:2002hc}.
Our results have been confirmed with the 
{\sc HELAC-1LOOP} implementation of the OPP
method~\cite{Ossola:2006us,vanHameren:2009dr,Czakon:2009ss}
within the statistical Monte Carlo error of 0.2\%~\cite{Bevilacqua:2009zn}.

\section{Description of the calculation}
\label{se:calculation}%\refse{se:calculation}
In NLO QCD, hadronic $\Pt\bar\Pt\Pb\bar\Pb$ production involves 
the $2\to 4$ partonic channels
$q\bar q\to\Pt\bar\Pt\Pb\bar\Pb$ (\mbox{7 trees} and 188 loop diagrams)
and $\Pg\Pg\to\Pt\bar\Pt\Pb\bar\Pb$ (36 trees and 1003 loop diagrams).
The $2\to 5$ bremsstrahlung contributions
comprise the crossing-symmetric channels
$q\bar q\to\Pt\bar\Pt\Pb\bar\Pb \Pg$, $q\Pg\to\Pt\bar\Pt\Pb\bar\Pb q$, 
and
$\Pg\bar q\to\Pt\bar\Pt\Pb\bar\Pb \bar q$
(64 diagrams each),
and the partonic process
$\Pg\Pg\to\Pt\bar\Pt\Pb\bar\Pb \Pg$ (341 diagrams). 
Each contribution has been 
worked out twice and independently, resulting in two completely
independent computer codes.
The treatment of the $q\bar q$- and gluon-induced
reactions are described in~\citere{Bredenstein:2008zb} and~\citere{Bredenstein:2010rs}, respectively. 
Here we mainly focus on the virtual corrections 
in the gg channel.

Feynman diagrams are generated with two independent version of 
{\sc FeynArts}~\cite{Kublbeck:1990xc,Hahn:2000kx} and handled
with two in-house {\sc  Mathematica} programs 
that perform algebraic manipulations and
generate {\sc Fortran77} code fully automatically.
One of the two programs
relies on {\sc FormCalc}~\cite{Hahn:1998yk} for preliminary algebraic
manipulations.
The interference of the one-loop and LO matrix elements,
summed over colours and helicities, is computed 
on a diagram-by-diagram basis,
\beqar
\sum_{\mathrm{col}}
\sum_{\mathrm{hel}}
\M^{(\onel)} \left(\M^{(\LO)}\right)^*
&=&
\sum_{\Gamma}
\left[\sum_{\mathrm{col}}
\sum_{\mathrm{hel}}
\M^{(\Gamma)} \left(\M^{(\LO)}\right)^*
\right].
\eeqar
Individual loop diagrams $(\Gamma)$ are evaluated
by separate numerical routines and summed explicitly.
The cost related to the large number of diagrams is compensated
by the possibility to perform colour sums very efficiently thanks to 
colour factorization [see \refeq{colofactb}].
Individual (sub)diagrams
consist of a single colour-stripped
amplitude ${\cal A}^{(\Gamma)}$ multiplied by
a simple colour structure,\footnote{
More precisely, each quartic gluon coupling 
generates three independent colour structures
that are handled as separate subdiagrams.
However, most diagrams do not involve quartic couplings,
and their colour structure factorizes completely.}
which is easily reduced to a compact colour basis ${\{\cal
C}_{k}\}$. 
The LO amplitude is handled as a vector in colour space,
and colour sums are encoded once and for all in
a colour-interference matrix $I_{kl}$, 
\beqar\label{colofactb}%\refeq{colofactb}
\M^{(\Gamma)}
&=&
\A^{(\Gamma)}
\left(\sum_k c_k^{(\Gamma)}{\cal C}_{k}\right),
\qquad
\M^{(\LO)}
=
\sum_{l}
\M_{l}^{(\LO)}
{\cal C}_{l},
\qquad
I_{kl}=
\sum_{\mathrm{col}} 
{\cal C}_k
{\cal C}^*_{l}.
\eeqar
These ingredients yield colour-summed results 
by means of {\em a single evaluation}  of the colour-stripped
amplitude $\A^{(\Gamma)}$ of each (sub)diagram.
Tensor integrals with $N$ propagators and $P$ Lorentz indices
are expressed in terms of totally symmetric covariant structures
$\{g\dots g p\dots p\}^{\mu_1\dots\mu_P}_{j_1\dots j_P}$ involving
external momenta $p_1,\dots,p_{N-1}$ and
$g^{\mu\nu}$ in $D=4-2\varepsilon$ dimensions~\citere{Denner:2005nn},
\beq
\frac{(2\pi\mu)^{4-D}}{\ri\pi^{2}}\int \rd^{D}q\,
\frac{q^{\mu_1}\dots q^{\mu_P}}
{\prod_{i=0}^{N-1}\left[(q+p_{i})^2-m_i^2+\ri 0\right]}
=
\sum_{j_1,\dots,j_P=0}^{N-1} 
{T^{N}_{j_1\dots j_P}}\; 
{\{g\dots g p\dots p\}^{\mu_1\dots\mu_P}_{j_1\dots j_P}}.
\eeq
The gg channel involves tensor integrals up to rank $P=4$.
The coefficients  $T^{N}_{j_1,\dots,j_P}$
are related to scalar integrals by means of
{\em numerical algorithms} that avoid instabilities
from inverse Gram determinants and other spurious 
singularities~\cite{Denner:2002ii,Denner:2005nn}.
The tensor rank and the number of propagators of 
integrals with $N>4$ are simultaneously
reduced without introducing inverse Gram determinants.  
Tensor integrals with $N=4,3$ are handled 
with the Passarino--Veltman algorithm as long
as no small Gram determinant appears in the reduction.  
Otherwise, 
expansions about the limit of vanishing Gram determinants
and possibly other kinematical determinants are applied~\cite{Denner:2005nn}.
The reduction is strongly boosted by a cache system
that recycles tensor
integrals among diagrams with common subtopologies.  The
$\Pg\Pg$ channel involves about
350 scalar integrals, which require
$10\,$ms CPU time per phase-space point.\footnote{All CPU times refer to a 
3 GHz Intel Xeon processor.}
The calculation of all scalar and tensor integrals
with and without cache system takes 
$40\,$ms and $200\,$ms, respectively. 
Rational terms arising from ultraviolet (UV) poles of 
tensor integrals with $D$-dependent coefficients
are automatically extracted by means of
a catalogue of residues $R^{N}_{j_1\dots j_P}$,
\beqar\label{rationalb}%\refeq{rationalb}
f(D) T^{N}_{j_1\dots j_P}
&=&
f(D)\left(\hat{T}^{N}_{j_1\dots j_P}
+
\frac{R^{N}_{j_1\dots j_P}}{\epsilon_{\mathrm{UV}}}\right)=
f(4) T^{N}_{j_1\dots j_P}
-2 f'(4) R^{N}_{j_1\dots j_P}.
\eeqar
Rational terms resulting from infrared poles 
must be taken into account only in 
wave-function renormalization factors,
since they cancel in truncated one-loop amplitudes~\cite{Bredenstein:2008zb}.

\newcommand{\QG}[5]{
\gamma^{\mu_{#1}}
\gamma^{\mu_{#2}}
\gamma^{\mu_{#3}}
\gamma^{\mu_{#4}}
\gamma^{\mu_{#5}}
}
\newcommand{\TG}[3]{
\gamma^{\mu_{#1}}
\gamma^{\mu_{#2}}
\gamma^{\mu_{#3}}
}
\newcommand{\SG}[1]{
\gamma^{\mu_{#1}}
}
\newcommand{\GTE}[2]{
g^{\mu_{#1}\mu_{#2}}
}

The helicity-dependent parts of all diagrams are reduced 
to a common basis of so-called Standard Matrix Elements (SMEs)
of the form
\beq\label{SMEs}%\refeq{SMEs}
{\hat\M}_m =
Q_m^{\mu_1\mu_2\rho_1\dots \rho_l}
\varepsilon_{\mu_1}(p_1)
\varepsilon_{\mu_2}(p_2)
\left[\bar v(p_3) \gamma_{\rho_1}\dots \gamma_{\rho_k}u(p_4)\right]
\left[\bar v(p_5) \gamma_{\rho_{k+1}}\dots
\gamma_{\rho_l}u(p_6)\right],\nn\nl
\eeq
where $Q_m^{\mu_1\mu_2\rho_1\dots \rho_l}$ are
combinations of metric tensors and external momenta.
The colour-stripped
part of each loop 
diagram [see \refeq{colofactb}]
yields a linear combination of SMEs and tensor integrals,
\beqar\label{SMEdecomp}%\refeq{SMEdecomp}
\A^{(\Gamma)}
&=&
\sum_m {\cal F}^{(\Gamma)}_m {\hat \M}_m,
\qquad
{\cal F}^{(\Gamma)}_m = \sum_{P} 
\sum_{j_1,\ldots,j_P=0}^{N-1} 
{\cal K}^{(\Gamma)}_{m;j_1\dots j_P}
{T^{N}_{j_1\dots j_P}} 
\;+\; \mbox{rational parts}.
\eeqar
The use of  SMEs  enables very efficient helicity summations.
Helicity and colour sums are encoded in the
interference of SMEs ${\hat\M}_m$ and colour structures ${\cal C}_k$
with the LO amplitude,
\beqar
M_{km}
&=&
\sum_{\mathrm{col}}
\sum_{\mathrm{hel}}
{\hat\M}_m 
{\cal C}_k
\left(\M^{(\LO)}\right)^*
=
\sum_{l}
I_{kl}
\sum_{\mathrm{hel}}
 {\hat\M}_m 
\left(\M^{(\LO)}_{l}\right)^*.
\eeqar
This matrix, which must be computed only once per phase-space point,
links the
colour/helicity-independent form factors
${\cal F}^{(\Gamma)}_m$ of each diagram
to its colour/helicity-summed
contribution
\beqar\label{colhelsum}%
\sum_{\mathrm{col}}
\sum_{\mathrm{hel}}
\M^{(\Gamma)} \left(\M^{(\LO)}\right)^*
&=&
\sum_m {\cal F}_m^{(\Gamma)}
\left(\sum_{k}
c_k^{(\Gamma)} 
M_{km}\right).
\eeqar
The SME-reduction 
starts with process-independent $D$-dimensional relations
such as momentum conservation, Dirac algebra, 
transversality, and gauge-fixing conditions for the
gluon-polarization vectors.
Once rational terms are extracted, we further 
reduce SMEs with two alternative algorithms in four dimensions.
The first algorithm splits each fermion chain 
into two contributions,
$
u(p_{i})
=
\sum_{\lambda=\pm}\omega_\lambda
u(p_{i}),
$  via insertion of chiral projectors 
$\omega_\pm=(1\pm\gamma^5)/2$.
This permits to employ various relations of 
type
$
{
\gamma^\mu}
{\gamma^\alpha
\gamma^\beta}
\omega_\pm
\otimes
{
\gamma_{\mu}}
=
{
\gamma^\mu}
\omega_\pm
\otimes
\left(
\gamma_{\mu}
\gamma^\beta
\gamma^\alpha
\omega_\pm
+
\gamma^\alpha
\gamma^\beta
\gamma_{\mu}
\omega_\mp
\right)$,
which %are based on Chisholm's identity and 
connect Dirac matrices of %thatbelong to 
different fermion chains~\cite{Bredenstein:2008zb,Denner:2005fg}.
In this way a rich variety of non-trivial identities
are obtained that reduce the full amplitude to 502 SMEs~\cite{Bredenstein:2010rs}.
Besides this procedure, which  depends on 
process-specific aspects, we implemented
a simple process-independent reduction
based on a  four-dimensional
identity of type
$\QG{1}{2}{3}{4}{5}=\GTE{1}{2}\TG{3}{4}{5}-\GTE{1}{2}\GTE{3}{4}\SG{5} + \mbox{perm.}$,
which eliminates
spinor chains with more than three Dirac matrices without
introducing $\gamma_5$~\cite{Bredenstein:2010rs}.
This leads to 970 SMEs. 
In spite of the factor-two difference in the number of SMEs, we found that
the numerical codes based on the two different reductions have the same---and 
remarkably high---CPU speed: about 180 ms per phase-space point.
This unexpected result means that the obtained CPU performance, at
least for this process, does not depend on process-dependent optimisations.

\newcommand{\qparbar}{\raisebox{.6em}{\tiny $(-)$}\hspace{-.83em}q}

To handle the real corrections we employed
the dipole subtraction method~\cite{Catani:2002hc,Frederix:2008hu}. 
The $2\to 5$ matrix elements 
were generated with {\sc Madgraph}~\cite{Alwall:2007st}
and checked against analytic calculations
and in-house code based on off-shell recursions.
%The phase-space integration is performed with
%multi-channel Monte Carlo generators.
More details are given in~\citere{Bredenstein:2010rs}.

\section{Predictions for the LHC}
\label{se:numres}%\refse{se:numres}

We present results for $\Pp\Pp\to\Pt\bar\Pt\Pb\bar\Pb+X$ at
$\sqrt{s}=14\TeV$ for $\Mt=172.6\GeV$
and $m_\Pb=0$.
Collinear final-state partons
are recombined into
jets with 
%rapidity--azimuthal-angle separation
$\sqrt{\Delta\phi^2+\Delta y^2}>0.4$
using a \mbox{$k_{\rT}$-algorithm}.
We impose the cuts 
$p_{\rT,\Pb}>20\GeV$ and $|y_\Pb|<2.5$
on b jets and use the CTEQ6 %~\cite{Pumplin:2002vw} 
set of PDFs with $N_{\mathrm{F}}=5$ active flavours and 
$\Lambda_5^{\overline{\mathrm{MS}}}=226\MeV$.  
Further details are given in~\citere{Bredenstein:2010rs}.

In all recent ATLAS studies of $\Pt\bar\Pt\PH (\PH\to\Pb\bar\Pb)$~\cite{Aad:2009wy} the signal and its
$\Pt\bar\Pt\Pb\bar\Pb$ background are simulated 
by setting the
renormalization and factorization scales equal to half the threshold
energy, $E_{\mathrm{thr}}=2 \Mt+m_{\Pb\bar\Pb}$.
Being proportional to $\alpha^4_{\mathrm{s}}$, these LO 
predictions are extremely sensitive to the scale choice,
and in~\citere{Bredenstein:2009aj} we found that
at $\mu_\mathrm{R,F}=E_{\mathrm{thr}}/2$ the 
NLO corrections to $\Pp\Pp\to\Pt\bar\Pt\Pb\bar\Pb$ 
are close to a factor of two.
This enhancement is due to the fact that 
$\Pp\Pp\to\Pt\bar\Pt\Pb\bar\Pb$
is a multi-scale process
involving various scales well below $E_{\mathrm{thr}}/2$.
In particular, the cross section is saturated by b quarks with
$p_{\rT,\Pb}\ll \Mt$~\cite{Bredenstein:2010rs}.  
In order to avoid large logarithms 
we advocate the use of the dynamical scale
\beq\label{centralscale}%\refeq{centralscale}
\mu^2_0=\Mt\sqrt{p_{\rT,\Pb}p_{\rT,\bar\Pb}},
\eeq
which improves the perturbative convergence and minimises NLO effects
in the shape of distributions~\cite{Bredenstein:2010rs}.
%%%%%%%%%%%%%%%%%%%%%%%%%%%%%%%%%%%%%%%%%%%%%%%%%%%%%%%%%%%%
% setupI
%%%%%%%%%%%%%%%%%%%%%%%%%%%%%%%%%%%%%%%%%%%%%%%%%%%%%%%%%%%%%
\begin{figure}
\includegraphics[bb= 95 445 280 655, width=.40\textwidth]
{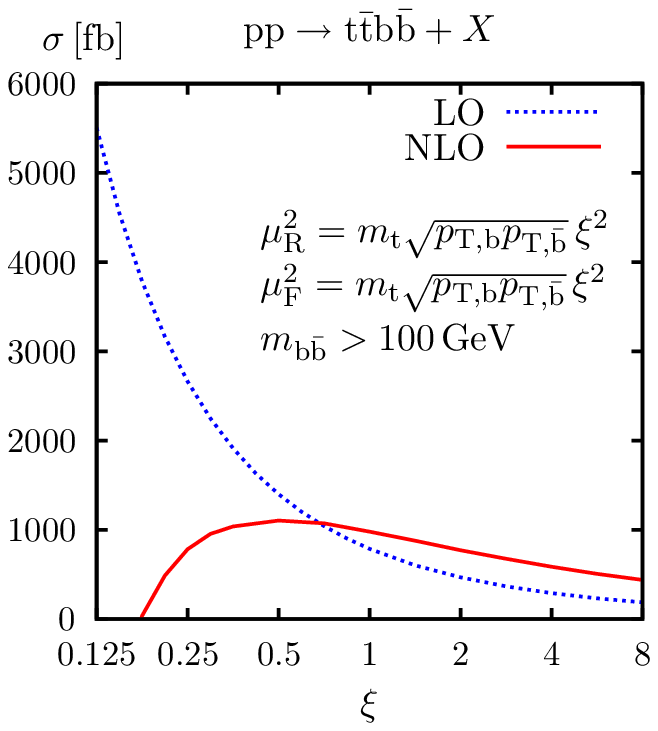}
\hfill
\includegraphics[bb= 95 445 280 655, width=.40\textwidth]
{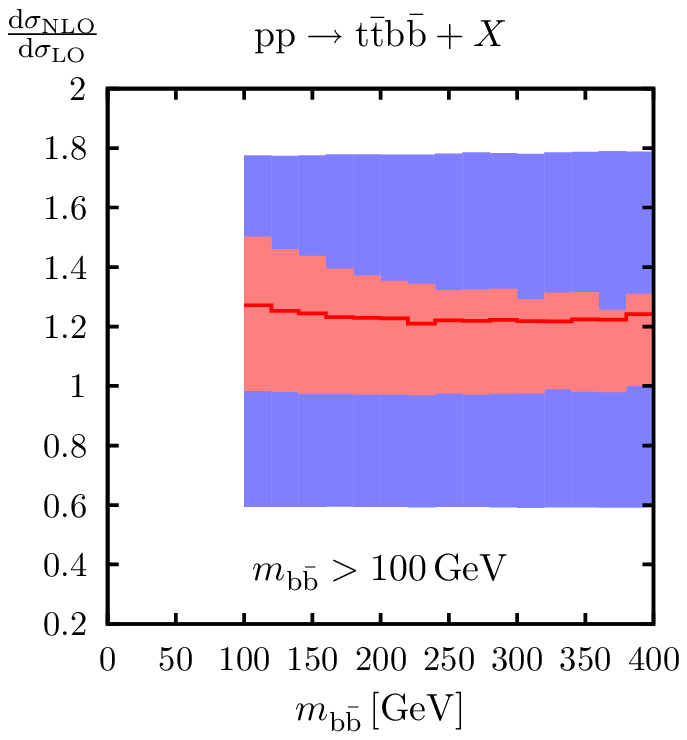}
\vspace*{-.8em}
\caption{$\Pp\Pp\to\Pt\bar\Pt\Pb\bar\Pb+X$ cross section
at the LHC for standard cuts and $m_{\Pb\bar\Pb}>100\GeV$: 
scale dependence of the LO and NLO cross section
(left plot) and relative NLO corrections
to the $m_{\Pb\bar\Pb}$ distribution (right plot).
}
\label{fig:setupI}
\end{figure}
%%%%%%%%%%%%%%%%%%%%%%%%%%%%%%%%%%%%%
In \reffi{fig:setupI} we show results for the 
kinematic region $m_{\Pb\bar\Pb}>100\GeV$,
which is relevant for  ATLAS/CMS studies of 
$\Pt\bar\Pt\PH(\PH\to\Pb\bar\Pb)$.
The left plot displays the dependence of the LO and NLO cross sections 
with respect
to scale variations $\mu_{\mathrm{R}}=\mu_{\mathrm{F}}=\xi \mu_{0}$.
At the central scale we obtain
$\sigma_{\mathrm{LO}}=786.3(2)\fb$ and
$\sigma_{\mathrm{NLO}}=978(3)\fb$.
This NLO result is $2.18$ times larger as compared to 
the LO cross section based on the ATLAS scale choice~\cite{Bredenstein:2010rs}.
The scale choice \refeq{centralscale} reduces the
$K$ factor to $1.24$, and the NLO (LO) uncertainty 
corresponding to factor-two scale variations
amounts to 21\% (78\%). The improvement with
respect to~\citere{Bredenstein:2009aj}
%, where we had a 33\% NLO uncertainty, 
is evident also from the stability of the NLO 
curve in \reffi{fig:setupI}a.
The right plot in \reffi{fig:setupI} displays LO (blue) and NLO (red)
scale-dependent predictions for the $m_{\Pb\bar\Pb}$
distribution. The results are normalized to the LO distribution at
$\mu_{\mathrm{R,F}}=\mu_0$ and the bands correspond to factor-two scale
variations.
The NLO predictions
perfectly fit within the LO band
and exhibit little kinematic dependence. 
Inspecting various other distributions
we found similarly small NLO corrections 
to their shapes. 
%%%%%%%%%%%%%%%%%%%%%%%%%%%%%%%%%%%%%%%%%%%%%%%%%%%%%%%%%%%%
% setupII
%%%%%%%%%%%%%%%%%%%%%%%%%%%%%%%%%%%%%%%%%%%%%%%%%%%%%%%%%%%%%
\begin{figure}
\includegraphics[bb= 95 445 280 655, width=.40\textwidth]
{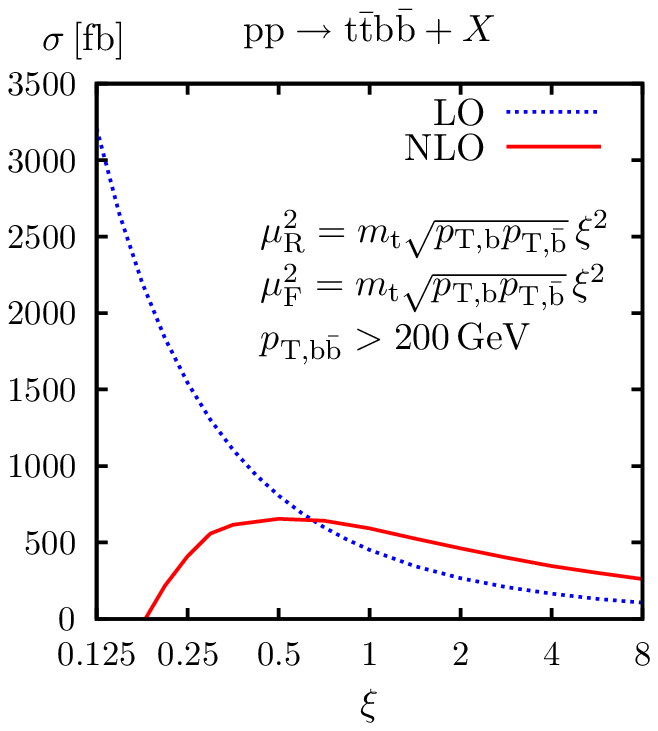}
\hfill
\includegraphics[bb= 95 445 280 655, width=.40\textwidth]
{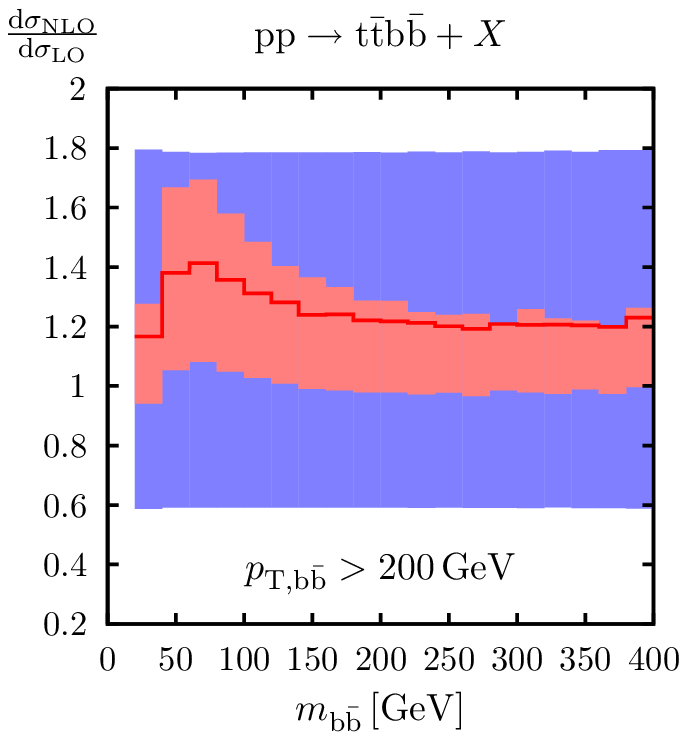}
\vspace*{-.8em}
\caption{$\Pp\Pp\to\Pt\bar\Pt\Pb\bar\Pb+X$ cross section
at the LHC for standard cuts and $p_{\rT,\Pb\bar\Pb}>200\GeV$: 
scale dependence of the LO and NLO cross section
(left plot) and relative NLO corrections
to the $m_{\Pb\bar\Pb}$ distribution (right plot).
}
\label{fig:setupII}
\end{figure}
%\vspace*{4em}
%%%%%%%%%%%%%%%%%%%%%%%%%%%%%%%%%%%%%%%%%%%%%%
In \reffi{fig:setupII} we show similar plots for the kinematic region
$p_{\rT,\Pb\bar\Pb}>200\GeV$, which permits to increase the
separation between the Higgs signal and its 
$\Pt\bar\Pt\Pb\bar\Pb$ background~\cite{Plehn:2009rk}.
At the central scale we obtain
$\sigma_{\mathrm{LO}}=451.8(2)\fb$ and
$\sigma_{\mathrm{NLO}}=592(4)\fb$. 
The $K$ factor (1.31), and the LO (79\%) and NLO (22\%)
scale dependence behave similarly as for $m_{\Pb\bar\Pb}>100\GeV$.
But in this case the NLO corrections are rather sensitive
to $m_{\Pb\bar\Pb}$. In the physically interesting
region of $m_{\Pb\bar\Pb}\sim 100\GeV$, 
the shape of the
$m_{\Pb\bar\Pb}$ distribution is distorted by about 20\%.
This effect tends to mimic a Higgs signal and should be carefully
taken into account in the $\Pt\bar\Pt\PH(\PH\to\Pb\bar\Pb)$ analysis.

\section{Conclusions}
\label{se:conclusion}%\refse{se:conclusion}

The observation of the $\Pt\bar\Pt\PH(\PH\to\Pb\bar\Pb)$ signal
and the direct measurement of the top-quark Yukawa coupling at the LHC
 require a very  precise description of the $\Pp\Pp\to
\Pt\bar\Pt\Pb\bar\Pb$ irreducible background.
The NLO QCD corrections to $\Pt\bar\Pt\Pb\bar\Pb$ production reveal that the
scale choice adopted in previous LO studies 
underestimates this cross section by a factor of two.
We advocate the use of a dynamical scale that stabilizes the perturbative
predictions reducing the $K$ factor to about $1.2$.
In presence of standard cuts NLO effects feature small 
kinematic dependence. But in the regime of highly boosted Higgs bosons 
we observe significant distortions in the shape of distributions.

The calculation is based on process-independent algebraic techniques, which
reduce loop diagrams to standard colour/helicity structures, and numerically
stable tensor-reduction algorithms. The very high numerical stability and
CPU efficiency of this approach are very encouraging in view of future NLO
calculations for multi-particle processes.

\end{document}